\documentclass[%
 reprint,
preprintnumbers,
 amsmath,amssymb,
 aps,
]{revtex4-2}

\usepackage{graphicx}
\usepackage{dcolumn}
\usepackage{bm}
\usepackage{hyperref}
\usepackage[mathlines]{lineno}
\usepackage{lineno}
\usepackage{color}
\newcommand{\red}{\color{black}}
\newcommand{\black}{\color{black}}

\begin{document}

\preprint{INR-TH-2022-XXX}

\title{General constraints on sources of high-energy cosmic rays from the interaction losses}

\author{Simon Sotirov}

 \email{sotirov.sa19@physics.msu.ru}
\affiliation{Physics Department, M.V. Lomonosov Moscow State University, 1-2 Leninskie Gory,  Moscow 119991, Russia}
\affiliation{Institute for Nuclear
Research of the Russian Academy of Sciences, 60th October Anniversary
Prospect 7a, Moscow 117312, Russia}

\begin{abstract}
Sources of high-energy cosmic rays are presently unknown, but can be constrained in various ways. Some of these constraints can be graphically presented on the so-called Hillas diagram. Previous versions of this diagram determined the range of geometrical sizes and magnetic fields of potential astrophysical accelerators, taking into account geometrical criteria and radiation losses. In this work, we update the Hillas diagram for protons, taking into account the losses associated with the $p\gamma$ \red and photopair \black interactions, relating the allowed regions to the source electromagnetic luminosity. The strongest constraints are obtained for bright compact sources such as the central regions of active galactic nuclei. 
\end{abstract}

\maketitle

\section{\label{sec:level1}introduction}
To date, despite many studies on high-energy cosmic rays and astrophysical neutrinos, their sources are still not unambiguously determined \cite{Palladino:2020jol, Kachelriess:2019oqu}. Therefore, it is important to impose constraints on the sources physical properties in order to narrow down the area of their search. The best known condition is the Hillas criterion \cite{Hillas:1984ijl} or geometry criterion: a particle must not leave the accelerator region until it has gained the required energy. It is usually assumed that the particle is held by the magnetic field and accelerated by the electric field. The next bound is due to the fact that when a charged particle moves with acceleration, it certainly radiates energy. This radiation hinders the accumulation of energy, so it must also be taken into account when constraining potential sources of acceleration of ultra-high-energy particles. These constraints can be represented graphically in the form of a diagram in which the magnetic field and source size are plotted, the Hillas diagram. Lines that meet these criteria limit the allowable range of parameters of astrophysical sources that are capable of accelerating particles to the corresponding energies. Detailed calculations and corresponding diagrams are presented, e.g. in \cite{1,Medvedev, Protheroe:2004rt, Aharonian:2002we}.

The purpose of this work is to further narrow down the class of potential sources of cosmic rays and, as a consequence, of high-energy neutrinos. To do this, one can take into account the proton energy loss due to interaction with source photons. Significant photon concentrations are achieved at relatively low energies. $\Delta^+$ production and multipion production are taken as the main contributions to the photohadronic cross section \cite{2}. Also this work takes into account the Bethe-Heitler process. Calculation of energy losses for interactions, is carried out assuming that the proton is accelerated by the diffuse acceleration (for example, shock waves) \cite{Rieger:2004jz}. For particular assumptions about radiation fields, these losses are presented in the Hillas diagram, thereby narrowing down the area of potential sources that are capable of accelerating protons to the corresponding energies.
 \section{energy of protons and energy losses}

\subsection{Radiation losses}
Radiation losses can be composed of the synchrotron and curvature radiations \cite{LL}. In the ultrarelativistic regime, the later is generally negligible. The synchrotron losses are dominant for any generic field configuration; however, in a very specific regime when particle velocity, a magnetic field and an electric field are parallel: $\mathbf{v} ||\mathbf{B} ||\mathbf{E}$ they vanish, and the losses
are then determined by the curvature radiation. The synchrotron losses for a proton moving in an astrophysical source with magnetic field $B$ is given by \cite{Medvedev}:
\begin{equation}
    -\frac{dE_{rad}}{dt} = \frac{2}{3} \frac{q^4}{m_p^4} E^2B^2 = D \left (\frac{E}{\text{eV}} \right)^2 \left ( \frac{B}{\text{G}} \right )^2 \text{eV/s},
\end{equation}
where $q$ - proton charge, $m_p$ is a proton mass, $E$ - proton energy, $D = 1.5\cdot10^{-29}$.
\subsection{Photohadronic interaction}
\subsubsection{Interactions}

The most important contribution to the total cross section is the creation and subsequent decay of $\Delta ^+$ \cite{2, Troitsky:2021nvu}:

\begin{equation}
p + \gamma \to \Delta^+ \to 
 \begin{cases}
   n + \pi^+ \\
   p + \pi^0 .
 \end{cases}
\end{equation}
 Consideration of the kinematics of the two-particle decay of $\Delta ^+$ and pions decay gives us that the energy carried away by the neutrino $E_{\nu}$ is about 20 times less than the energy of the incident proton $E_p$. In addition, by describing the resonance condition: $ E_pE_{\gamma t} = m_{\Delta}^2 $, one can estimate the energy of the initial target photon $E_{\gamma t}$:
\begin{equation}
    E_{\gamma t} \simeq \frac{m_{\Delta}^2}{20E_{\nu}},
\end{equation}
where $m_{\Delta}$ is $\Delta^+$ mass.
It should be noted that this ratio is written in the reference frame of the source, so the observed values may differ. Doppler boosting and redshift must be taken into account for this.

The second contribution is multipion production, which can be seen as a statistical process \cite{2}. In this case, the proton loses about $60\%$ of its initial energy.

The total $p\gamma$ interaction cross section $\sigma_{p\gamma}$ can be estimated as the sum of these two channels, as a result of which an analytical approximation can be used \cite{2}:

\begin{equation}
    \sigma_{p\gamma}(\epsilon_r) K_{p\gamma }(\epsilon_r) = 70H(\epsilon_r-\epsilon_{\text{thr}}) \ \mu \text{b},
    \label{cross section p gamma}
\end{equation}
where $\epsilon_r = \gamma_p \epsilon(1-\beta_p \mu)$ is energy invariant of the interaction, which is equal to the photon energy in the proton’s rest frame (in units of an electron mass), where $\mu = \cos\theta$, $\theta$ is the angle between the colliding proton and photon momenta. $\gamma_p=(1 - \beta_p^2)^{-1/2}$ is proton Lorentz factor in units of a proton mass, $ \epsilon$ is energy of the target photons in units of an electron mass, $K_{p\gamma}$ is the inelasticity of the collision, \red $\epsilon_{\text{thr}} = 390$ \black and $H$ is the Heaviside step function. 

\subsubsection{Interaction losses}

Here the proton energy losses during interaction with the photons of the medium are calculated. The inverse of the photohadronic energy-loss timescale for high energy protons is given by (in the particle physics $c=1$ units) \cite{2}:
\begin{equation}
    \label{pg los} t_{\gamma p}^{-1} \cong \frac{1}{2\gamma_{p}^{2}}\int_0^\infty \,\frac{n_{\text{ph}} (\epsilon)}{\epsilon^2}\mathrm{d}\epsilon \!\int_{0}^{2\gamma_{p}\epsilon}\!\epsilon_r\sigma_{p \gamma }(\epsilon_r) K_{p\gamma }(\epsilon_r) \,\mathrm{d}\epsilon_r,
\end{equation}
where  $n_{\text{ph}}(\epsilon)$ is a spectral number density of target photons.

We will consider a spherical source with radius $R$ and power law spectrum $n_{\text{ph}} = \beta \epsilon^{-\alpha}$. \red The coefficient $\beta$ can be expressed in terms of $L_{\geq \omega}$ (for $\alpha > 2$) or $\tilde{L}_{\geq \omega}$ (for $\alpha \leq 2$), where $L_{\geq \omega}$ is the source bolometric luminosity at a photon energy greater than $\omega =  m_e \epsilon_{\text{thr}}/2\gamma_p$ and $\tilde{L}_{\geq \omega}$ is the same but for photon energies greater than $\omega$ and less than $\Omega$. Where $\Omega$ is the high energy cutoff of a source spectrum and $m_e$ is an electron mass. By using (\ref{cross section p gamma}) and performing the integral over the interval from $\omega$ to $+\infty$ (or from $\omega$ to $\Omega$ in the case of a hard spectrum) in (\ref{pg los}) \black the expression for the proton energy losses due to $p\gamma$ interactions can be obtained in the form:
\begin{equation}
   \frac{dE}{dt} = 
    \begin{cases}
        \displaystyle  -K_1 \left(\frac{E}{\text{eV}}\right)^2 \left( \frac{R}{\text{kpc}} \right)^{-2} \left( \frac{L_{\geq \omega}}{\text{eV/s}} \right) \text{eV/s}, & \alpha > 2 \\[3mm]
        \displaystyle  -K_2 \left(\frac{E}{\text{eV}}\right)^2 \left( \frac{R}{\text{kpc}} \right)^{-2} \left( \frac{\tilde{L}_{\geq \omega}}{\text{eV/s}} \right) \text{eV/s}, &  \alpha = 2 \\[3mm]
        \displaystyle -K_3 \left(\frac{E}{\text{eV}}\right)^\alpha \left( \frac{R}{\text{kpc}} \right)^{-2} \left( \frac{\tilde{L}_{\geq \omega}}{\text{eV/s}} \right) \text{eV/s}, & \alpha < 2 
        
    \end{cases}
    \label{Losses1}  
\end{equation}
where

\begin{eqnarray}
    K_1 = (\alpha\!-\!2)/(\alpha^{2}\!-\!1)\cdot 10^{-89}
\end{eqnarray}

\begin{eqnarray}
    K_2 = 5 \cdot 10 ^{-90} \left( \ln \frac{\Omega}{\omega} \right)^{-1}
    \label{hard1}
\end{eqnarray}

\begin{equation}
    K_3 = \frac{2-\alpha}{\alpha^2 - 1} 6^{\alpha} \cdot 10 ^ {-12\alpha - 68} \Omega^{\alpha - 2}
    \label{hard2}
\end{equation} 
\black \subsection{Photopair process}
In addition to $p\gamma$ interaction there is important for high energy protons process \cite{Zatsepin:1966jv, Stecker:1968uc, Stecker:1969fw}: $p + \gamma \to p + e^- + e^+$. The expression for energy loss is obtained in \cite{Blumenthal:1970nn}. In a proton case we have:
\begin{equation}
   \label{bhh los} \frac{dE}{dt} = - \alpha_f r_e^2 m_e \int_2^{\infty }d\epsilon\ \ n_{\text{ph}} \left(\frac{\epsilon}{2\gamma_p}\right)\frac{\varphi(\epsilon)}{\epsilon^2} ,
\end{equation}
where $\alpha_f$ is the fine structure constant, $r_e$ is the classical electron radius. We take the following fit with the relative error $< 1.5 \times 10^{-3}$ \cite{1992ApJ...400..181C}:
\begin{equation}
     \varphi(\epsilon) \cong \frac{\pi}{12} \frac{(\epsilon - 2)^4}{1 +  \sum_{i=1}^4 c_i(\epsilon - 2)^i}  \ \ \text{for} \ \ 2 < \epsilon < 25,
\end{equation}
where $c_1 = 0.8048, c_2 = 0.1459, c_3 = 1.137 \cdot 10^{-3} , c_4 = -3.879\cdot 10^{-6}$, and

\begin{equation}
    \varphi(\epsilon) \cong \frac{\epsilon \sum_{i=0}^3 d_i\ln^i \epsilon}{1 - \sum_{i=1}^3 f_i \epsilon^{-1}} , \ \ \text{for} > 25,
\end{equation}
where $d_0=-86.07, d_1 = 50.96, d_2 \cong -14.45, d_3 \cong 2.67, f_1 \cong 2.91, f_2 = 78.35, f_3 = 1837$.
In the power-law case $n_{\text{ph}}(\epsilon) = \beta \epsilon^{-\alpha}$ we obtain:

\begin{equation}
    \frac{dE}{dt} =
    \begin{cases}
        \displaystyle -C_1 \left(\frac{E}{\text{eV}}\right)^{2} \left( \frac{R}{\text{kpc}} \right)^{-2} \left( \frac{L_{\geq \omega}}{\text{eV/s}} \right) \text{eV/s}, & \alpha > 2,\\[3mm]
        \displaystyle -C_2 \left(\frac{E}{\text{eV}}\right)^{2} \left( \frac{R}{\text{kpc}} \right)^{-2} \left( \frac{\tilde{L}_{\geq \omega}}{\text{eV/s}} \right) \text{eV/s}, & \alpha = 2,\\[3mm]
        \displaystyle -C_3 \left(\frac{E}{\text{eV}}\right)^{\alpha} \left( \frac{R}{\text{kpc}} \right)^{-2} \left( \frac{\tilde{L}_{\geq \omega}}{\text{eV/s}} \right) \text{eV/s}, & \alpha < 2,
    \end{cases}
    \label{Losses2}
\end{equation}

where 

\begin{equation}
    C_1 = 10^{-89} (\alpha - 2) \epsilon_{\text{thr}}^{\alpha - 2} I(2)
\end{equation}

\begin{equation}
    C_2 = 2 \cdot 10 ^{-89} \left( \ln \frac{\Omega}{\omega} \right)^{-1} I(2)
    \label{hard3}
\end{equation}

\begin{equation}
    C_3 = 10^{-71} (2-\alpha) \left( \frac{\Omega}{m_e} \right)^{\alpha-2} \left ( \frac{2 \cdot 1\text{eV}}{m_p} \right )^{\alpha} I(\alpha)
    \label{hard4}
\end{equation}

\begin{equation}
    I(\alpha) =  \int_2^{\infty} \frac{\epsilon^{-\alpha} \varphi(\epsilon)}{\epsilon^2} d\epsilon
\end{equation} 
\subsection{Acceleration model and constraints on the sources}

By analogy with \cite{Medvedev}, only the diffuse mechanism acceleration model is considered here. In the most common scenarios of such acceleration, the particle moves inside the accelerator and from time to time receives a portion of energy as a result of interaction with a shock. Since this regime assumes a disordered configuration of fields, the energy losses and radiation will be due only to synchrotron radiation \cite{1}.

Let us consider the propagation of a particle in a magnetized source medium. The particle acquires energy due to repeated scattering on a shock, after which the particle moves for long distances along an approximately Larmor orbit, while radiating energy and interacting with low-energy photons of the medium until it again receives a push from a next shock, then the process repeats. As we will see, the maximum energy that a particle has when it leaves the source depends weakly on the energy that the particle received directly at the front of a shock, and for the most part is determined by losses.

Let us consider a particle having an initial energy (i.e., at the front of the shock wave) $E_0$ propagating through a region of a size $~R$ filled with a magnetic field $B$ and leave a source. As the particle is moving in the medium, its energy decreases due to interactions and radiations:
\begin{equation}
    \frac{dE}{dl} = - DE^2B^2 - K_iE^{\xi_i}R^{-2}L_i  - C_i E^{\xi_i} R^{-2} L_i,
\end{equation}
where $l$ is the distance along the particle trajectory, $i = 1,2,3$. $\xi_1 = \xi_2 = 2, \ \ \xi_3 = \alpha$, $L_1 = L_2 = L_{\geq \omega}$, $L_3 =  \tilde{L}_{\geq \omega}$, see (\ref{Losses1}) and (\ref{Losses2}). Assuming that the magnetic field $B(l)$ is approximately constant over distances of the order $R$. Particle escaping energy $E'$ is determined by the integral equation:
\begin{equation}
    \int_{E_0}^{ E'} \frac{dE}{DE^2B^2 + R^{-2}L_i(K_iE^{\xi_i} + C_iE^{\xi_i})} = -R,
\end{equation}
the maximum escaping energy $E_{\text{cr}}$ is obtained in limit $E_0 \to \infty$. 
\begin{equation}
    \int_{E_{\text{cr}}}^{ \infty} \frac{dE}{DE^2B^2 + R^{-2}L_i(K_iE^{\xi_i} + C_i E^{\xi_i})} = R.   
\end{equation}
Therefore, the region where protons can be accelerated to energies of $10^{\lambda}$ eV is determined by the inequality: $E_{cr} \geq 10^{\lambda} \text{ eV}$. Let's graphically depict this inequality (augmented Hillas diagram).

It is also necessary to take into account the geometric criterion (Hillas criterion): the Larmor radius of the particle must not exceed the size of the accelerator, otherwise the particle will leave the accelerator before it acquires sufficient energy. It can be represented as an inequality \cite{Hillas:1984ijl}:

\begin{equation}
    E \leq A \left( \frac{B}{\text{G}} \right) \left( \frac{R}{\text{kpc}} \right) \ \text{eV}, 
\end{equation}
where $A = 9.25 \cdot 10^{23} $. All these constraints for the diffuse acceleration mode are presented graphically in Fig.~\ref{fig:wide}.
\section{DISCUSSION AND IMPLICATIONS}
\subsubsection{UHENCRs in central regions of AGNs}
\begin{figure*}
\includegraphics{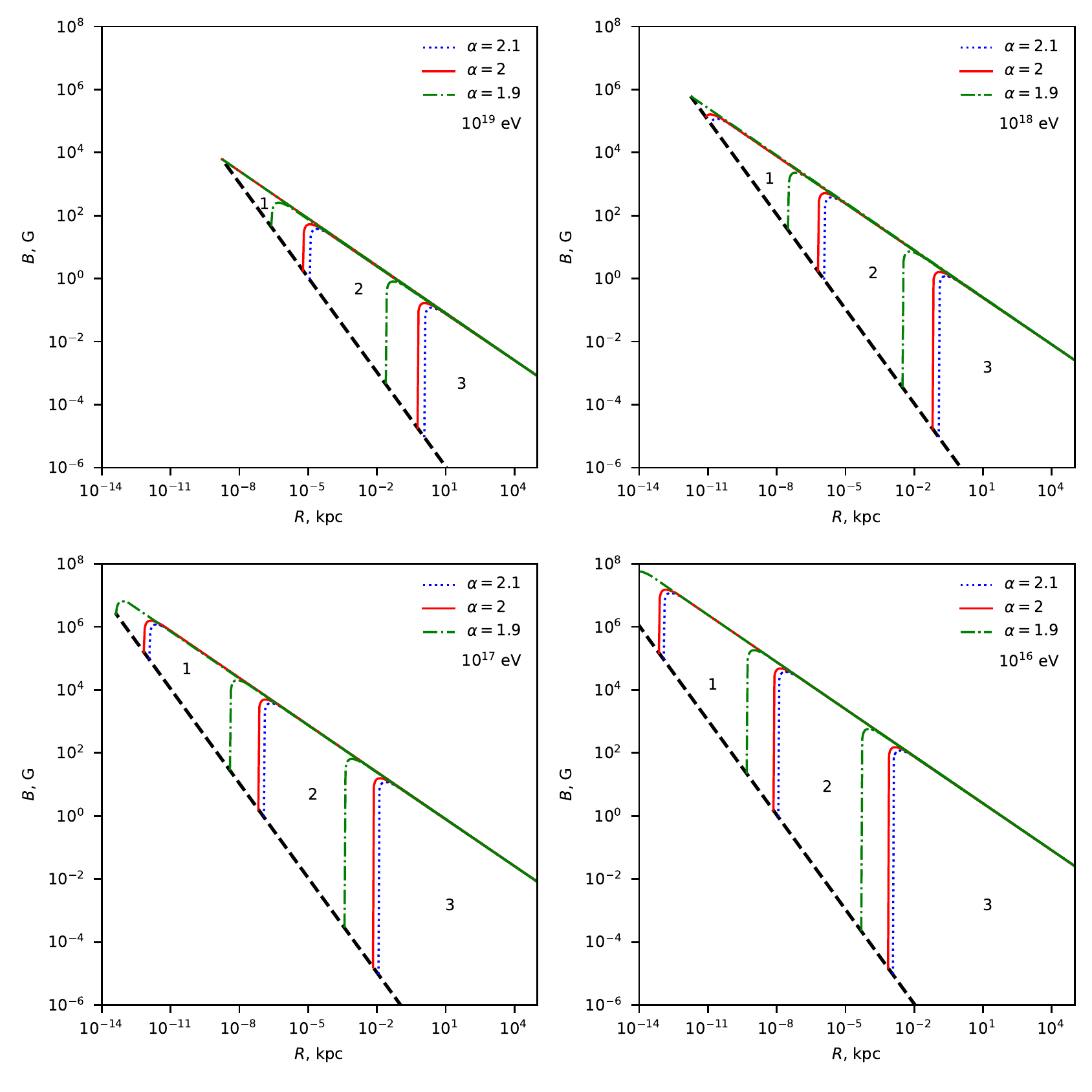}
\caption{\label{fig:wide} Magnetic field - source size diagram for proton acceleration, assuming power-law radiation field. The black dotted line is the Hillas geometric criterion. The dotted blue lines, solid red lines and dot-dashed lines are the limitations associated with losses due to radiation in a magnetic field and due to interaction with source photons for $\alpha = 2.1$, $\alpha = 2$ and $\alpha=1.9$ respectively. The lines limit the range of parameters that the source can have in order to accelerate protons to a given energy. In this case, each graph corresponds to a certain energy: $10^{16} \text{ eV}, 10^{17} \text{ eV}, 10^{18} \text{ eV},10^{19} \text{ eV}$. In the case of $10^{16}$ eV, the region 1+2+3 corresponds to the region of parameters (R,B) that a source with luminosity $L_{\geq 0.9 \text{ keV}} = 10^{38} \text{ erg s}^{-1}$ can have, in order to accelerate protons up to $10^{16}$ eV. Region 2+3 is similar, but for luminosity $L_{\geq 0.9\text{ keV}} = 10^{43} \text{ erg s}^{-1}$. Region 3 corresponds to luminosity $L_{\geq 0.9\text{ keV}} = 10^{48} \text{ erg s}^{-1}$. In the case of $10^{17}$ eV 1+2+3 corresponds to $L_{\geq 90\text{ eV}} = 10^{38} \text{ erg s}^{-1} $ , 2+3 corresponds to $L_{\geq 90\text{ eV}} = 10^{43} \text{ erg s}^{-1} $, Region 3 for $L_{\geq 90\text{ eV}} = 10^{48} \text{ erg s}^{-1} $. In the case of $10^{18}$ eV 1+2+3 corresponds to $L_{\geq 9\text{ eV}} = 10^{38} \text{ erg s}^{-1} $ , 2+3 corresponds to $L_{\geq 9\text{ eV}} = 10^{43} \text{ erg s}^{-1} $, Region 3 for $L_{\geq 9\text{ eV}} = 10^{48} \text{ erg s}^{-1} $. And finally, in the case of $10^{19}$ eV 1+2+3 corresponds to $L_{\geq 0.9\text{ eV}} = 10^{38} \text{ erg s}^{-1} $ , 2+3 corresponds to $L_{\geq 0.9\text{ eV}} = 10^{43} \text{ erg s}^{-1} $, Region 3 for $L_{\geq 0.9\text{ eV}} = 10^{48} \text{ erg s}^{-1} $. For spectra with $\alpha = 2$ and $\alpha=1.9$ it is assumed $\Omega=10^{15}$ eV.} 
\end{figure*}

With the help of the obtained results, the Hillas diagram was updated, supplemented by the constraints associated with the photohadronic and photopair interactions of protons with source photons. As a result, we have one more source parameter - luminosity. As can be seen from Fig.~\ref{fig:wide}, lower luminosity corresponds to a larger allowable range of parameters for accelerating particles to the corresponding energy. Therefore, one can try to impose constraints on compact and at the same time bright sources. Such, for example, are the central regions of the nuclei of active galaxies (AGNs).

Let us determine the location of the central regions of the AGNs on the Hillas diagram. The size of the potential region of AGNs acceleration is of the order of several gravitational radii $R_s$.
\begin{equation}
    R \sim 5R_s \approx 5 \times 10^{-8} \ \frac{M_{\text{BH}}}{10^8M_{\odot}} \text{ kpc ,}
    \label{eq:Size}
\end{equation}
where $M_{\text{BH}}$ is the black hole mass. $M_{\text{BH}}$ varies from $10^6 M_{\odot}$ for ordinary galaxies to $10^{10}M_{\odot}$ for powerful radio galaxies and quasars.

The magnetic field near the black hole's horizon $B_{\text{BH}}$ is highly dependent on the black hole's mass. A conservative estimate is obtained, for example, in \cite{Shakura}.
\begin{equation}
    B_{{\text{BH}}} \sim 10^8 \left(\frac{M_{\text{BH}}}{M_{\odot}}\right)^{-0.5} \text{ G}.
    \label{eq:Magn}
\end{equation}
Real values of $B_{\text{BH}}$ are 1-2 orders of magnitude lower, which will be taken into account in the diagram. 

\begin{figure*}
\includegraphics{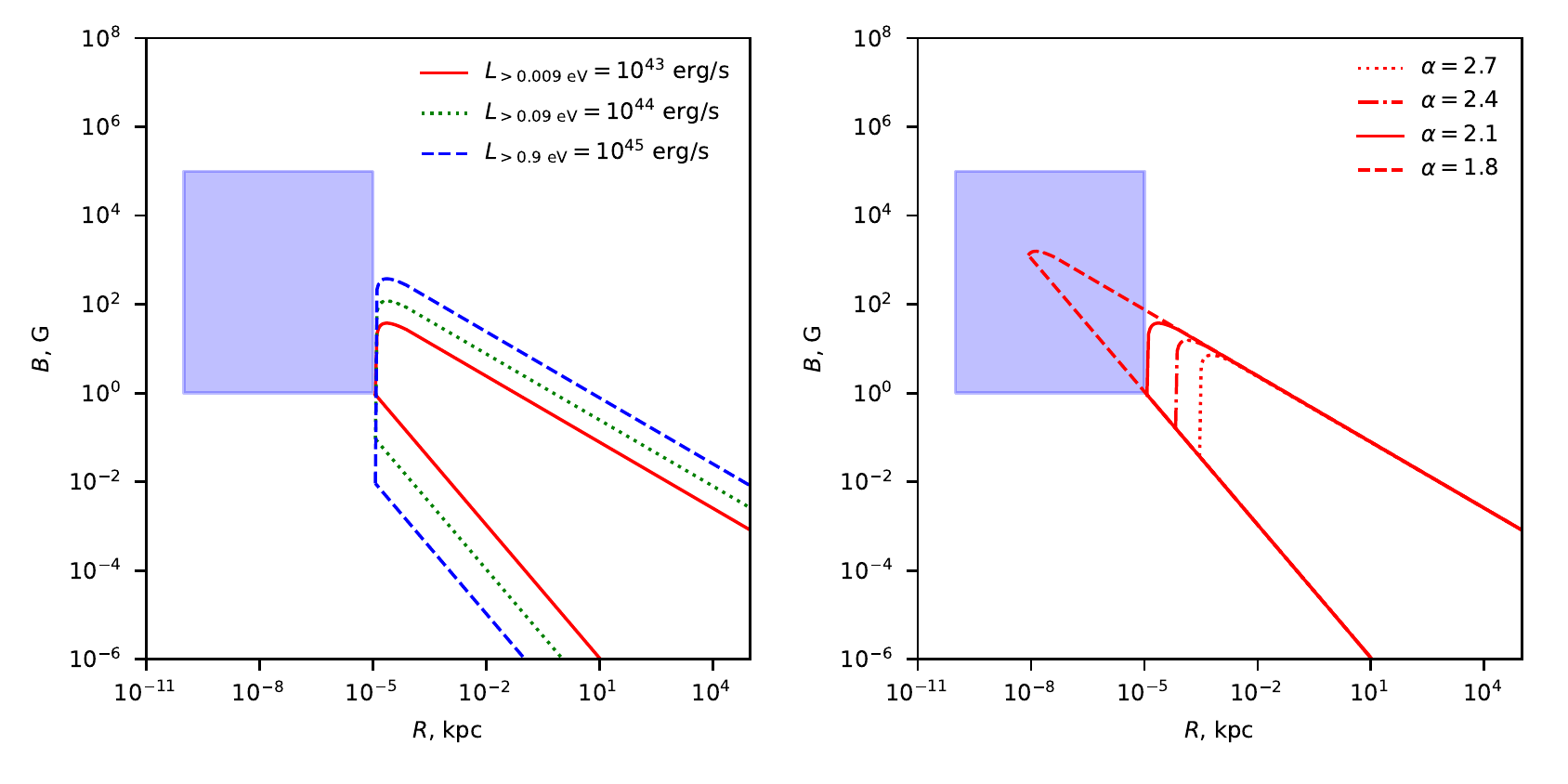}
\caption{\label{fig:2} Magnetic field $B$ - source size $R$ diagrams. The blue rectangle is the area of parameters $(R,B)$ that central regions of AGNs can have. $\mathtt{Left}:$ the solid red line limits the range of parameters that sources with luminosities $L_{\geq 0.009 \text{ keV}} = 10^{43} \text{ erg s}^{-1}$ can have in order to accelerate protons up to $10^{19}$ eV. The dotted green line limits the range of parameters that sources with luminosities $L_{\geq 0.09 \text{ keV}} = 10^{44} \text{ erg s}^{-1}$ can have in order to accelerate protons up to $10^{18}$ eV. The solid red line limits the range of parameters that sources with luminosities $L_{\geq 0.9 \text{ keV}} = 10^{45} \text{ erg s}^{-1}$ can have in order to accelerate protons up to $10^{17}$ eV. $\mathtt{Right}:$ lines for different spectral indices $\alpha$ for $10^{19}$ eV and $L_{\geq 0.009 \text{ keV}} = 10^{43} \text{ erg s}^{-1}$. For spectrum with $\alpha=1.8$ it is assumed $\Omega = 10^{15}$ eV.}
\end{figure*}

Using expressions (\ref{eq:Size}) and (\ref{eq:Magn}), it is possible to graphically depict the area of parameters of the central regions of AGNs on the Hillas diagram, which is shown in Fig.~\ref{fig:2}. It can be seen from the figure that for some $\alpha$ the central regions of AGNs with luminosities $L_{\geq 0.009 \text{ eV}}$ greater than $10^{43} \text{ erg s}^{-1}$ cannot accelerate protons up to $10^{19}$ eV and higher in the diffuse acceleration mode. If the luminosity $L_{\geq 0.09 \text{ eV}}$ exceeds $10^{44} \text{ erg s}^{-1}$, then AGNs cannot accelerate protons up to $10^{18}$ eV and higher. Similarly, for luminosities $L_{\geq 0.9 \text{ eV}}$ greater than $10^{45} \text{ erg s}^{-1}$ , there is no way to accelerate up to $10^{17}$ eV. 

Throughout this paper, we work in the reference frame of the source photon field (where the photons are isotropic), and $L_{\geq \omega}$ (or $\tilde{L}_{\geq \omega}$) denotes the total 4$\pi$ luminosity, which is a Lorentz invariant. For particular sources, especially for relativistic jets of active galactic nuclei, the Doppler enhancement of the observable flux should be taken into account when using observational data to estimate $L_{\geq \omega}$ (or $\tilde{L}_{\geq \omega}$), see e.g. \cite{2}.

\begin{figure*}
\includegraphics{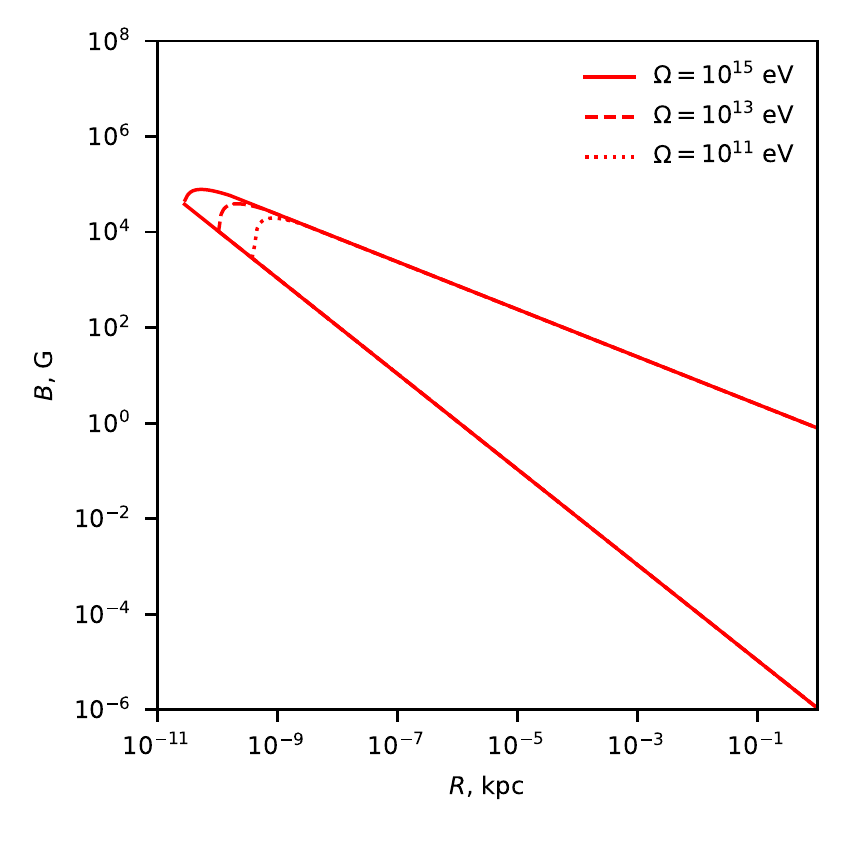}
\caption{\label{Omega} Magnetic field $B$ - source size $R$ diagram for different high-energy cutoff energies $\Omega$. The lines limit the range of parameters which sources with $\tilde{L}_{\geq \omega} = 10^{10}L_{\odot}$ and $\alpha=1.7$ can have in order to accelerate protons up to $10^{18}$ eV. }

\end{figure*}

It should be noted that all these constraints are made on the assumption that the source photon distribution is described by a power-law spectrum and diffuse acceleration mode is assumed. Also as can be seen from expressions \ref{hard1}, \ref{hard2}, \ref{hard3} and \ref{hard4}, in the case of a hard spectrum ($\alpha \leq 2$), the diagrams become sensitive to the high-energy cutoff $\Omega$, it is shown in Fig.~\ref{Omega}.

\subsubsection{Neutrino channel}  
Up to this point ultra high energy cosmic rays were the object of our attention. At the same time, high energy neutrinos can be produced in charged pions decays born in proton-photon collisions. These neutrinos can be observed in experiments, e.g. IceCube, ANTARES but so far it is impossible to say unequivocally from what sources they reach us. Therefore, it is also an important task to know in which source the neutrino production mechanism is effective and in which it is not. Having such information, one can try to narrow down the class of neutrino loud sources.

The ratio of proton and neutrino luminosities can serve as a quantitative characteristic of the efficiency of neutrino production $L_p/L_{\nu}$. If this ratio is small, then the neutrino production mechanism is more efficient; otherwise, fewer protons interact with photons and, consequently, fewer neutrinos are produced.

Let us consider the central regions of active galaxies as sources; for simplicity, we consider them spherically symmetrical with radius $R$ and photon luminosity $L$. As in the previous case, we take the spectral photon density in the form of a power law, where, for definiteness, the spectral index $\alpha = 2.1$. It is assumed that protons are accelerated in the central regions of the AGNs and interact only with photons of the source in the $\Delta$-resonant approximation, as a result of which neutrinos are born in the same region, which carry away $\sim$ 1/20 of the initial proton energy.

\begin{figure}

\includegraphics{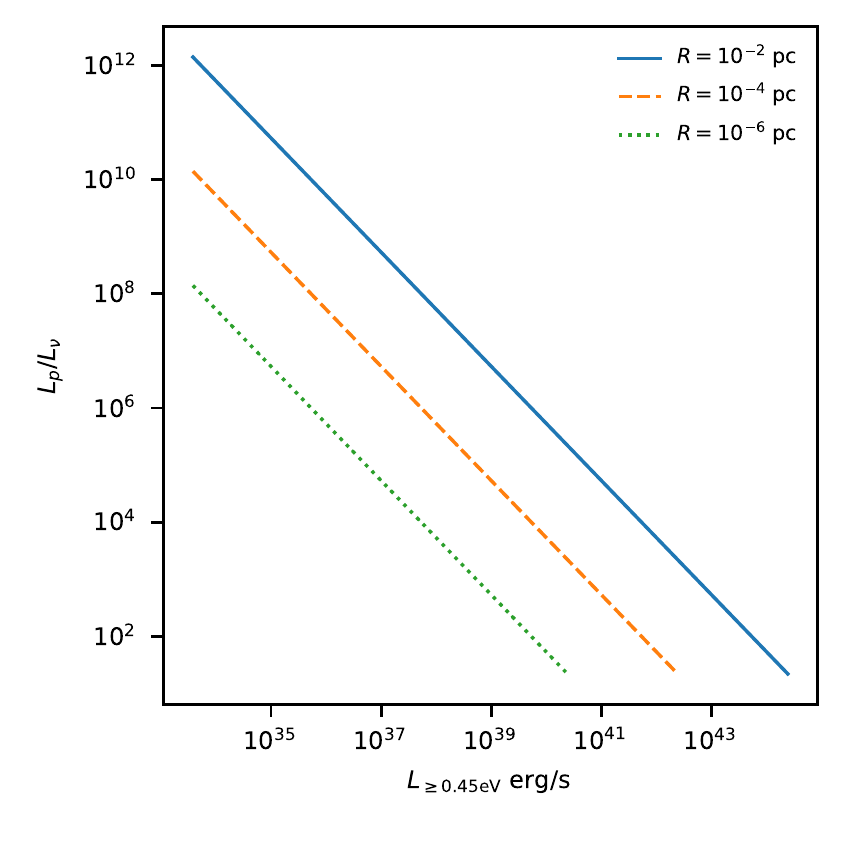}
\caption{\label{fig:3} The ratio of proton and neutrino luminosities depending on the bolometric photon luminosity of the source. The results are presented for various source sizes.}
    
\end{figure}

Astrophysical neutrinos with energies of order 10 TeV - 10 PeV are observed experimentally \cite{IceCube:2013low, ANTARES:2020srt, Katz:2006wv}. For simplicity, it is assumed that all neutrino sources are the central regions of the AGNs. Then, in the reference frame of the source, there should be protons with energies from 0.2 PeV to 200 PeV. Then, at the same time, for the production of neutrinos of such energies, source photons from the $\Delta$ resonance approximation with energies from 0.5 eV to 1.1 keV are required. Under these conditions, as an estimate, consider the following: $L_{\nu} \approx pL_p/20$, where $p \sim Rn\sigma_{\gamma p}$ is the probability of $p\gamma$ interacting at the source, $n$ is number of photons per unit volume in the energy range from 0.5 eV to 1.1 keV. The calculation results are shown in Fig.~\ref{fig:3}.

It is directly seen from the graph that compact and bright sources are the most neutrino loud. Moreover, for some photon luminosities $L$, neutrino luminosity $L_{\nu}$ becomes comparable with proton luminosity $L_{p}$. This means that the protons actively interact with the source photons, in other words, the mean free path is equal be the order of magnitude to $R$ or even shorter. In such sources, it is much harder for protons to acquire high energies. Thus, knowing the size of the assumed neutrino production region of the AGN, it is possible, using a simple relation, to estimate the photon luminosity at which the $p\gamma$ mechanism is effective. It is important to note that the energy losses due to proton radiation and the presence of electric and magnetic fields were not taken into account here.

\subsubsection{Photon channel}

As a result of the $p\gamma$ interaction, in addition to neutrinos, high-energy photons (from $\pi^0$ decays) are born, which carry away about 1/10 of energy of initial protons. It is also possible to look in this model for which photons the source is optically thin or thick. For this, it is necessary to estimate the optical depth of the source. Also, as in the previous paragraph, it is assumed that the source is spherically symmetric with radius $R$ and has a photon bolometric luminosity $L$ with a power-law spectrum. It is assumed that gamma rays mainly interact with the source photons, producing electron-positron pairs.

Photon-photon pair production is a threshold process, the threshold energy is given by:

\begin{equation}
    E_{tr} = \frac{2m_e^2c^4}{E_{\gamma}(1-\cos{\theta})},
\end{equation}
The cross section of the reaction $\sigma_{\gamma\gamma}$ can be integrated and averaged over the angle of interaction of photons, as a result of which the cross section can be written in an analytical form \cite{2004yt}:

\begin{equation}
\begin{gathered}
    \sigma_{\gamma\gamma} = \frac{3\sigma_T}{2s^2} \Biggl[ \left(s + \frac{1}{2}\ln s -\frac{1}{6} + \frac{1}{2s}\right)\ln(\sqrt{s} + \sqrt{s-1})- \\
    \left(s + \frac{4}{9} -\frac{1}{9s} \right)\sqrt{1 - \frac{1}{s}}\Biggr],
\end{gathered} 
\end{equation}
where $s = E_{\gamma}E_{t}/2m_e^2c^4$, $E_t$ - target photon energy, $\sigma_T$ is Thomson scattering cross section.
Expression for the optical depth \cite{2004yt}:
\begin{equation}
    \tau (E_{\gamma})= \int_{E_1}^{E_2} \sigma_{\gamma\gamma}(E_{\gamma}E)n_{ph}(E)RdE.
\end{equation}

\begin{figure}
    
    \includegraphics{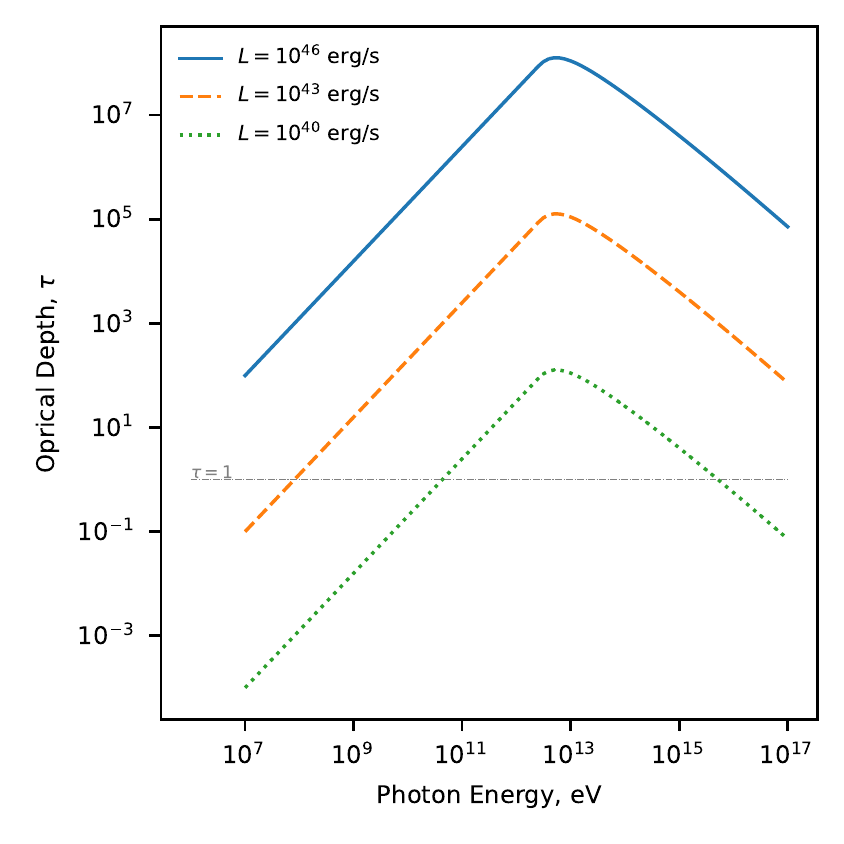}
    \caption{Gamma-ray optical depth in central regions of AGNs with size $R = 10^{-4}$ kpc. Each curve corresponds to bolometric luminosity $10^{40}$, erg/s, $10^{43}$ erg/s, $10^{46}$ erg/s from bottom to top, respectively, spectral index $\alpha = 2.1$.   }
    \label{fig:4}
\end{figure}

\begin{figure}
    
    \includegraphics{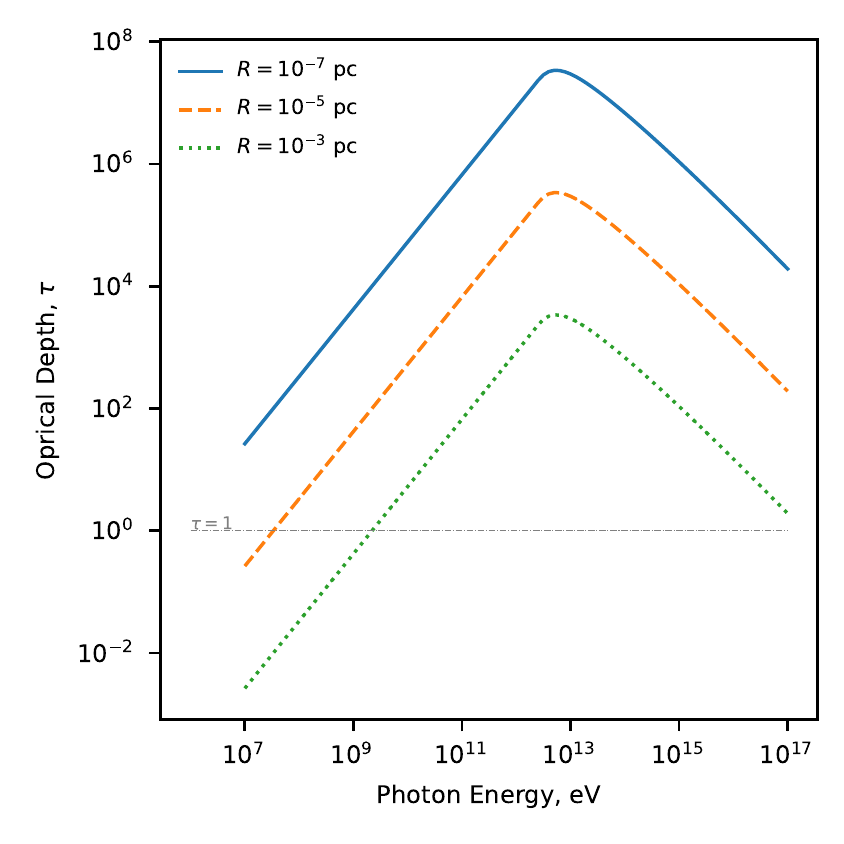}
    \caption{Gamma-ray optical depth in central regions of AGNs with luminosity $L = 10^{44}$ erg/s, spectral index $\alpha = 2.1$. Each curve corresponds to size $10^{-2}$, pc, $10^{-4}$ pc, $10^{-6}$ pc from bottom to top, respectively.   }
    \label{fig:5}
\end{figure}

\begin{figure}
    
    \includegraphics{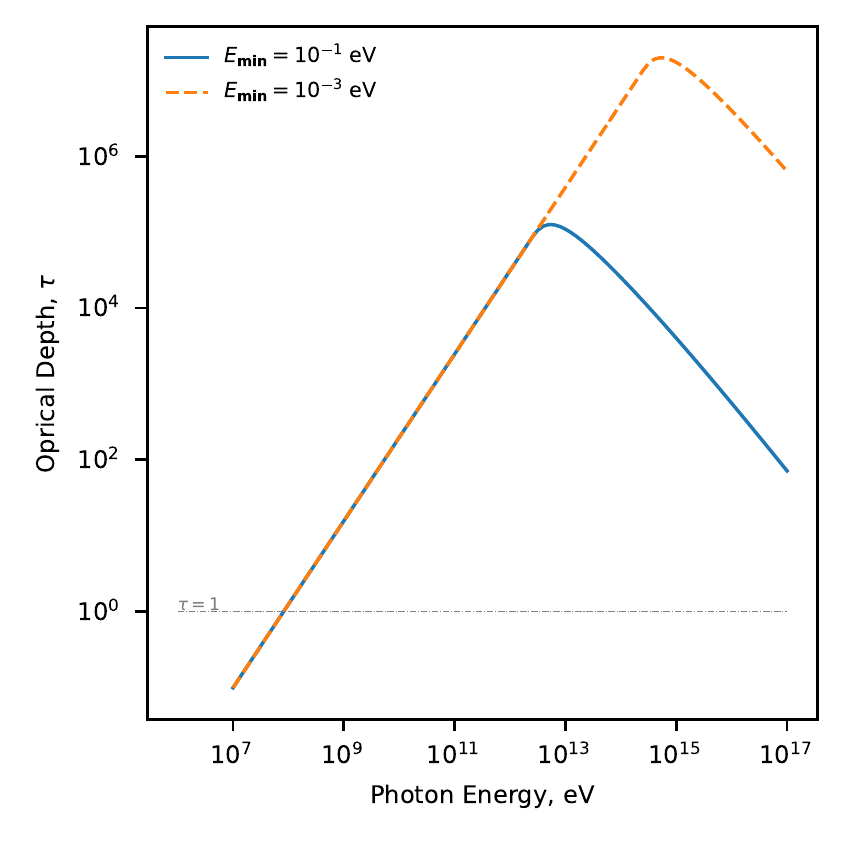}
    \caption{Gamma-ray optical depth in central region of AGNs for different initial energy $E_{\text{min}}$ with luminosity $L = 10^{43}$ erg/s , spectral index $\alpha = 2.1$ size $R=10^{-4}$ pc. Dashed curve corresponds to $E_{min} = 10^{-3}$ eV, solid line corresponds to $E_{\text{min}}=0.1$ eV.  }
    \label{fig:6}
\end{figure}

The calculation results are presented in Fig.~\ref{fig:4}. Within the framework of this model, it can be said that gamma rays resulting from the electromagnetic cascades started $p \gamma$ interactions from bright sources will not contribute to diffuse gamma radiation, which is observed by Fermi LAT. It can be seen from Fig.~\ref{fig:4} that for the central regions of AGNs with a size $ R=10^{-4}$ pc and luminosities exceeding $10^{45}$ erg/s, they do not contribute to diffuse gamma radiation. The dependence on the source size is shown at Fig.~\ref{fig:5}. All estimates are approximate and are given to understand the order of magnitude of physical quantities in this model. It is important to note that for the power-law spectrum the results depend on the minimal energy $E_{\text{min}}$. The results shown in graphs  Fig.~\ref{fig:3}, \ref{fig:4}, \ref{fig:5} are obtained for $E_{\text{min}} = 0.1$ eV. The result for the optical depth strongly depends on $E_{\text{min}}$, as shown in Fig.~\ref{fig:6}.

\subsubsection{Thermal radiation case}
It is also useful to consider the thermal spectrum, which can approximately describe central regions of AGNs of temperatures of $\sim$ (10-100) eV \cite{Shakura}. AGNs are potential sources of astophysical neutrinos. At the AGN center the super massive black hole (SMBH) surrounded by the accretion disc which emits thermal radiation resides. In this work it is assumed (see for example \cite{Stecker:1991vm, Kalashev:2014vya}) that proton acceleration occurs in the SMBH vicinity and then move along two jets perpendicular to the accretion disc. During propagation protons interact with the low energy photons coming from the accretion disc. As spectral number density on the distance $R_{\text{dist}}$  from disc along axis we take (see appendix):
\begin{equation}
    n_{\text{ph}} = x^2\frac{2\pi}{\lambda_{\text{C}}^3} \frac{\epsilon^2}{\exp(\epsilon/\Theta) - 1},
    \label{Thermal density}
\end{equation}
where $\lambda_{\text{C}}^3$ is the electron Compton wavelength, $\Theta \!=T/m_e$ is the dimensionless temperature of the radiation field, $x = R_{\text{AD}}/R_{\text{dist}}$. $R_{\text{AD}}$ is the typical accretion disc size which can be fitted as \cite{Morgan:2010xf}:
\begin{equation}
    R_{\text{AD}} = 10^{15} \left( \frac{M}{10^8 M_{\odot}} \right) \text{cm}. 
\end{equation}
Energy losses can be obtained using (\ref{cross section p gamma}), (\ref{pg los}) and (\ref{bhh los}):
\begin{eqnarray}
    \frac{dE}{dt} = - Q_1 \Theta^3 \varkappa(E) \left(\frac{E}{\text{eV}} \right) - Q_2 \xi(E) \left(\frac{E}{\text{eV}} \right)^{-2}\nonumber\\
     - D  \left(\frac{E}{\text{eV}} \right)^2 \left(\frac{B}{\text{G}} \right)^2 \ \ \text{eV}/ \text{s} ,\label{lossTh}
\end{eqnarray}
where $Q_1 = 3.82 \cdot 10 ^{12}$, $Q_2 = 6.5\cdot 10^{38}$, 
\begin{equation}
    \varkappa (E) = \int_{\tilde{\omega}}^{\infty} dy \ \frac{y^2 - \tilde{\omega}^2}{e^y - 1}, 
\end{equation}

\begin{equation}
    \xi (E) = \int_2^{\infty} d \epsilon \ \frac{\varphi(\epsilon)}{\exp(\epsilon m_p/2E\Theta) - 1},
\end{equation}
 and $\tilde{\omega} = m_e \epsilon_{\text{thr}}/2 \Theta \gamma_p$. 
 
\begin{figure*}
    \centering
    \includegraphics{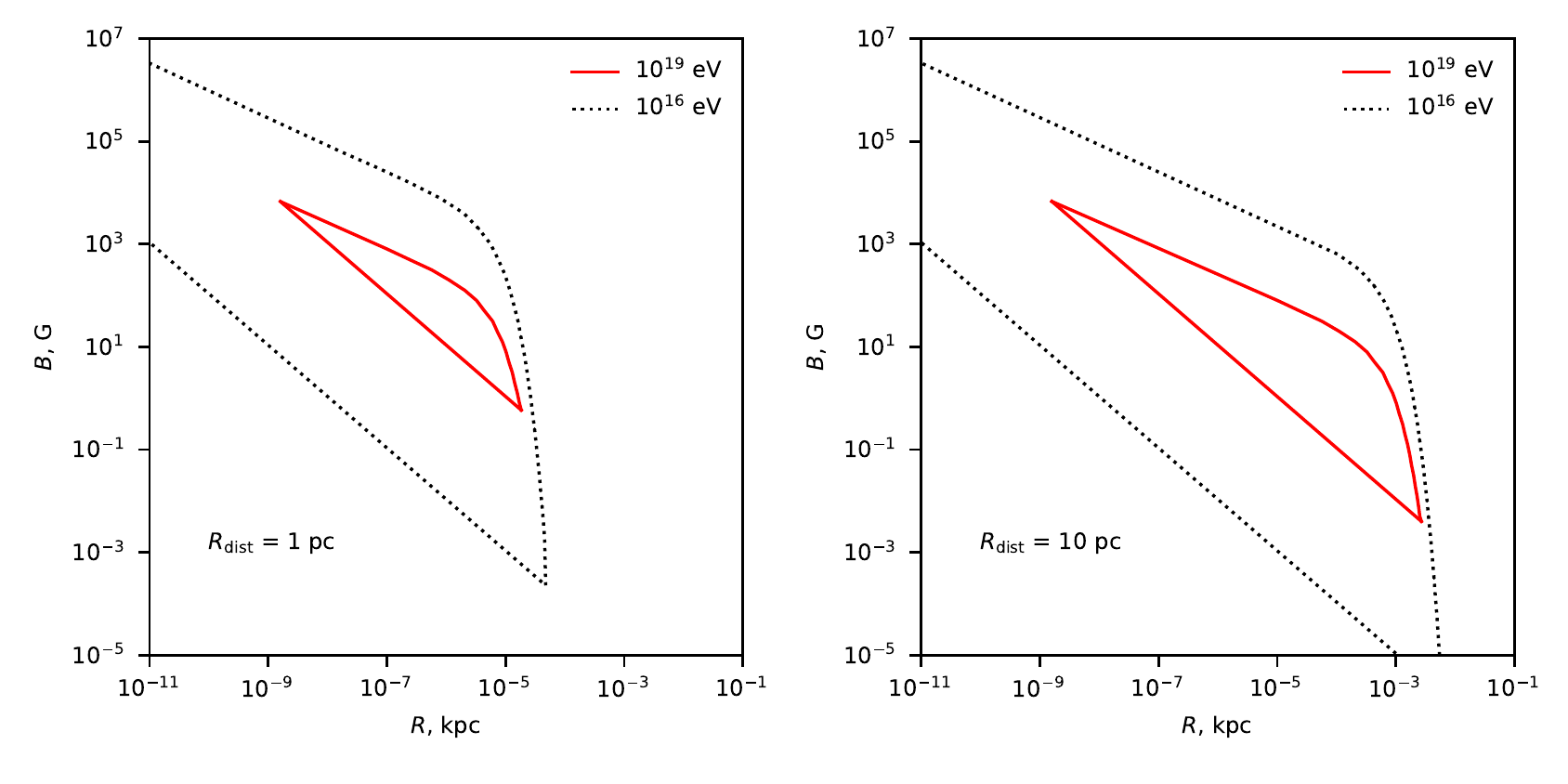}
    \caption{Magnetic field-size diagram. The red solid line limits the range of parameters at which the proton accelerated by a shock at the distance $R_{\text{dist}}$ from accretion disc can leave accelerator with energy $10^{19}$ eV. The same for the black dotted line for $10^{16}$ eV (see figure). It is assumed this that the SMBH has mass $10^8 M_{\odot}$ and accretion disc has temperature $\sim100$ eV. }
    \label{fig:Thermal}
\end{figure*}

\begin{figure*}
    \includegraphics{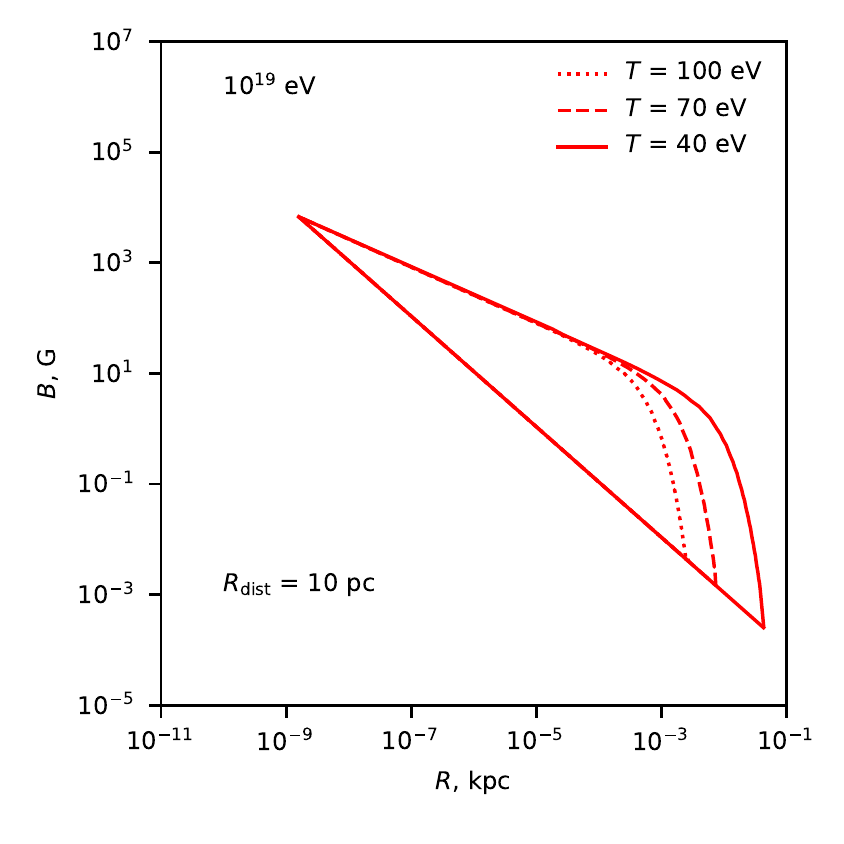}
    \caption{Magnetic field-size diagram. The red lines limits the range of parameters at which the proton accelerated by a shock at the distance $R_{\text{dist}} = 10$ pc from accretion disc can leave accelerator with energy $10^{19}$ eV for different temperatures 40 eV, 70 eV, 100 eV. It is assumed this that the SMBH has mass $10^8 M_{\odot}$.}
    \label{Thermal_T}
\end{figure*}

 Consider a proton propagating in a jet. Let it interact with a shock at distance $R_{\text{dist}}$ receive energy $E_0$ and leave the jet after passing distance $R$. Similarly, as was done for the power law spectrum in the limit $E_0 \to \infty$ we obtain Hillas diagram Fig.~\ref{fig:Thermal}. The lower the temperature, the larger the allowable area on the Hillas diagram, since with decreasing temperature the photon density (\ref{Thermal density}) decreases and the proton loses less energy for interaction. Diagrams for various temperatures are shown in Fig.~\ref{Thermal_T} In particular for energy $10^{19}$ eV, temperature 2.7 K and $x =1 $ the standard GZK cutoff $\sim 50$ Mpc can be turned out. 

\section{conclusions}

In contrast to previous researchers \cite{Medvedev,1}, in this work, a constraint was added to the Hillas diagram associated with the interaction of protons with source photons, as a result of which the region of allowable parameters becomes smaller, and the size of the region becomes parametrically dependent on luminosity or temperature. It can be seen from the obtained diagrams that compact and bright sources can be constrained most strongly, which was demonstrated with the central regions of the nuclei of active galaxies. In particular, a class of central regions of AGNs which can not accelerate protons up to $10^{19}$ eV, $10^{18}$ eV and $10^{17}$ eV was singled out assuming power-law radiation field and diffuse acceleration mechanism. In the case of thermal radiation field, it was also possible to significantly narrow down the search area for high-energy proton accelerators. As the next level of limitations, the fact that the size, magnetic field and luminosity of the central regions of AGNs are dependent on each other can serve.

The paper contains all the necessary formulas for constructing a diagram, so that the reader can build a diagram for their own needs, for example, if it is necessary to place a source with a given luminosity on the diagram and to understand whether the process of acceleration to a certain energy is possible in it.

\begin{acknowledgments}
The author is grateful to Sergey Troitsky for valuable remarks throughout the entire work and for careful reading of the manuscript. This work is supported by the RF Ministry of science and higher education under the contract 075-15-2020-778.
\end{acknowledgments}

\appendix*
\section{spectral photon density}
Consider the disc area element $r d\varphi dr$ at the distance $r$ from the SMBH. Contribution of such element to the photon density at point $R_{\text{dist}}$ along disc axis is 
\begin{equation}
    dn_{\text{ph}} = \frac{1}{2 \pi}\frac{r dr d \varphi}{r^2 + R_{\text{dist}}^2}n(\epsilon),
\end{equation}
where $n(\epsilon)$ is thermal photon density. Then photon density at the $R_{\text{dist}}$ is

\begin{equation}
    n_{\text{ph}} = \frac{1}{4}\ln\left( \frac{R_{\text{dist}}^2 + R_{\text{AD}}^2}{R_{\text{dist}}^2} \right) n(\epsilon) \approx  \frac{1}{4}\left(\frac{R_{\text{AD}}}{R_{\text{dist}}} \right)^2 n(\epsilon)
\end{equation}
The last equality follows from the assumption $R_{\text{dist}} \gg R_{\text{AD}}$. 
\bibliography{article}

\end{document}